\documentclass[12pt]{article}

\usepackage{scicite}

\usepackage{graphics}
\usepackage{color}
\usepackage{float}
\usepackage{graphicx}
\usepackage{amssymb}
\usepackage{overpic}
\usepackage{epstopdf}
\usepackage{wasysym}
\usepackage{bm}
\usepackage{graphicx}

\newenvironment{sciabstract}{%
\begin{quote} \bf}
{\end{quote}}

\newenvironment{methods}{%
    \section*{Methods}%
    \setlength{\parskip}{0pt}%
    }{}

\makeatletter
\def\@cite#1#2{\textsuperscript{{#1\if@tempswa , #2\fi}}}
\makeatother

\newcommand{\onlinecite}[1]{\hspace{-1 ex} \nocite{#1}\citenum{#1}}

\def\STF{$\kappa$-[(BEDT\--TTF)$_{1-x}$\-(BEDT\--STF)$_{x}$]$_2$\-Cu$_2$(CN)$_3$}
\def\stf{$\kappa$-STF$_x$}

\newcounter{lastnote}

\renewcommand{\figurename}{{\bf{Figure}}}
\renewcommand{\tablename}{{\bf{Table}}}

\makeatletter
\makeatletter \renewcommand{\fnum@figure}{{\bf{\figurename~\thefigure}}}
\makeatletter \renewcommand{\fnum@table}{{\bf{\tablename~\thetable}}}
\makeatother

\title{Failed superconductivity in a Mott spin liquid material}

\author
{Yuxin Wang,$^{1,2}$ Vladimir Dobrosavljevi\'{c},$^{1,2}$ Eun Sang Choi,$^{1}$\\  Yohei Saito,$^{3,4}$ Atsushi Kawamoto,$^{4}$ Andrej Pustogow,$^{5}$\\ Martin Dressel,$^{3}$ Dragana Popovi\'{c}$^{1,2\ast}$\\
\\
\normalsize{$^{1}$National High Magnetic Field Laboratory, Florida State University,}\\
\normalsize{Tallahassee, Florida 32310, USA}\\
\normalsize{$^{2}$Department of Physics, Florida State University,}\\
\normalsize{Tallahassee, Florida 32306, USA}\\
\normalsize{$^{3}$1.~Physikalisches Institut, Universit\"{a}t Stuttgart,}\\
\normalsize{70569 Stuttgart, Germany}\\
\normalsize{$^{4}$Department of Physics, Hokkaido University,}\\
\normalsize{Sapporo 060-0810, Japan}\\
\normalsize{$^{5}$Institute of Solid State Physics, Vienna University of Technology,}\\
\normalsize{Vienna, Austria}\\
\normalsize{$^\ast$To whom correspondence should be addressed; E-mail: dragana@magnet.fsu.edu}
}

\date{}

\begin{document} 


\baselineskip24pt


\maketitle 


\begin{sciabstract}

A central challenge for understanding unconventional superconductivity in most strongly correlated electronic materials is their complicated band structure and presence of competing orders.  In contrast, quasi-two-dimensional organic spin liquids are single-band systems with superconductivity arising near the bandwidth-tuned Mott metal-insulator transition in the absence of other orders.  Here, we study chemically substituted $\bm{\kappa}$-organics in which superconducting fluctuations emerge in the phase coexistence region between the Mott insulator and the Fermi liquid.  Using magnetotransport and ac susceptibility measurements, we find that global superconductivity fails to set in as temperature $\bm{T\rightarrow 0}$.  Our results indicate instead the presence of superconducting domains embedded in the metallic percolating cluster that undergo a magnetic field-tuned quantum superconductor-to-metal phase transition.  Surprisingly, albeit consistent with the percolation picture, universal conductance fluctuations are seen at high fields in macroscopic samples.   The observed interplay of the intrinsic inhomogeneity and quantum phase fluctuations provides a new insight into failed superconductivity, a phenomenon seen in various conventional and unconventional superconductors, including cuprates.
 
\end{sciabstract}

Unconventional superconductivity\cite{Stewart2017} (SC) has been reported in a wide range of strongly correlated materials, including cuprates\cite{Keimer2015}, heavy fermion and organic systems\cite{Scalapino2012}, iron-based materials\cite{Fernandes2022}, magic-angle twisted graphene\cite{Cao2018}, and layered metallic kagome compounds\cite{Hossain2025}.  However, its precise physical origin has proved difficult to unravel in most systems because of their complicated electronic structure, often assuming a multi-band or multi-orbital character within a unit cell, as well as the presence of many competing energy scales.    Furthermore, in many materials such as cuprates, unconventional SC emerges near an insulating or magnetically ordered state, but understanding the interplay of the Mott physics with SC remains a long-standing problem.  In this regard, quasi-two-dimensional (quasi-2D) organic charge-transfer salts\cite{dressel2020advances} have attracted considerable attention, precisely because many of their properties can be understood in terms of the simplest single-band Hubbard model\cite{Powell2011, Pustogow2018, Menke2024}.  In particular, the so-called $\kappa$-organics, which can be tuned across the Mott metal-insulator transition (MIT) by varying the bandwidth\cite{dressel2020advances}, feature a single half-filled band near the Fermi energy and SC with typical transition temperatures $T_{\mathrm{c}}\sim 10$~K.  Remarkably, several materials in this family, those identified as spin liquids, do not display any sign of magnetic, charge, or structural orders accompanying the transition, while clear signatures of SC arise on the metallic side\cite{Powell2011, dressel2020advances}.  Thus these materials represent the simplest model systems to investigate strongly correlated phenomena and the related unconventional SC.  However, tuning across the Mott transition is usually done by applying external pressure, which makes many experiments difficult or unfeasible, especially at cryogenic temperatures.

\begin{figure}[!htbp]
\centering
\includegraphics[width=0.95\textwidth]{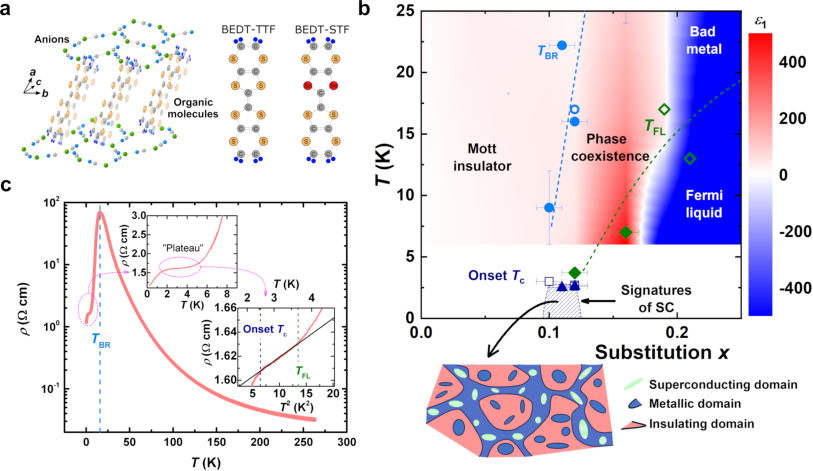}
\caption{\textbf{Emergence of superconductivity in $\bm{\kappa}$-STF$\bm{_x}$.}  \textbf{a}    The crystal structure contains dimers of the organic donor molecules (BEDT-TTF) and (BEDT-STF) forming layers in the $bc$-plane, which are separated by the Cu$_2$(CN)$_3$ anion sheets.  In (BEDT-STF), two sulfur atoms of the inner rings are replaced by selenium.  One electron per dimer is donated to the anion sheets; each dimer carries one hole with spin $1/2$, such that within the $bc$ plane the dimers are arranged in anisotropic triangles with a high degree of frustration (Supplementary Fig.~1a).   \textbf{b}  $T$--$x$ phase diagram at low $T$, where the Brinkman-Rice temperature $T_{\mathrm{BR}}$ separates the gapped Mott insulator from the phase coexistence region.  $T_{\mathrm{BR}}$ is determined from the peak in $\rho(T)$, as illustrated in Fig.~\ref{fig:Mixture}c (see also Supplementary Figs.~1b-d).  Near the boundary of the phase coexistence region, $T_{\mathrm{c}} < T_{\mathrm{FL}}$, where $T_{\mathrm{c}}$ indicates the onset of SCFs (Fig.~\ref{fig:Mixture}c lower inset).  Solid symbols: this work; open symbols from refs.~\onlinecite{Pustogow2021FL, Pustogow2021, Saito2021crystl}.  Color map: dielectric permittivity $\epsilon_1$ (ref.~\cite{Pustogow2021}) reveals a sharply defined insulator-metal phase-coexistence region, a mixture of metallic and insulating domains, around the first-order MIT.  Our work focuses on the behavior at even lower $T< T_{\mathrm{c}}$ (see hatched area) for which the sketch of the phase coexistence region summarizes some of the key conclusions of our work: the presence of superconducting domains embedded in the metallic percolating cluster, as $T\rightarrow 0$.  \textbf{c} $\rho(T)$ for sample S1 with a nominal $x=0.12$.  The low-$T$ regime, studied in this work, is marked by the oval and shown in the insets.  In this sample, $\rho(T)$ starts to deviate from the FL behavior at $T= T_{\mathrm{c}}=(2.6\pm 0.1)$~K, below which it exhibits a faster drop with decreasing $T$.
}
\label{fig:Mixture}
\end{figure}

Recent work\cite{Saito2021JMCC, Pustogow2021FL, Pustogow2021} has revealed, though, that most features of the bandwidth-tuned Mott transition in these materials can be alternatively induced by chemical substitution, which increases the electronic bandwidth $W$ while remaining at ambient pressure.  In particular, in charge-transfer salts \STF\, (abbreviated \stf), arguably an ideal realization of the single-band Hubbard model on a half-filled triangular lattice, partial substitution of S atoms in the organic (BEDT-TTF) donor molecules by Se results in Se-containing (BEDT-STF) with more extended orbitals and, hence, a  larger $W$ (Fig.~\ref{fig:Mixture}a and Supplementary Fig.~1).
%
%
Detailed studies, combining experiment and theory, have firmly established a number of issues concerning the normal phase, including the formation or destruction of a strongly correlated Fermi liquid (FL) on the metallic side\cite{Pustogow2021FL}, and the physical nature of the coexistence region around the Mott transition\cite{Pustogow2021}.  In addition, the onset of superconducting fluctuations (SCFs) was observed at $T_{\mathrm{c}}\sim 2-3$~K within the insulator-metal phase coexistence region\cite{Saito2021crystl}, but no bulk SC was found down to the lowest experimental $T\approx 1.4$~K.  Therefore, since dimers of (BEDT-TTF) and (BEDT-STF) in \stf\, are arranged in an almost ideal triangular lattice in the 2D planes (Fig.~\ref{fig:Mixture}a and Supplementary Fig.~1), making this material a well-known quantum spin liquid candidate\cite{Pustogow2023}, the important question about the interplay of unconventional SC and the Mott MIT in systems with a high degree of geometrical frustration remains open\cite{Menke2024}.  Moreover, while SC is generally probed by applying a magnetic field ($H$), understanding of quantum ($T=0$) phase transitions (QPTs) from a superconducting to a resistive state in 2D that occurs by tuning some parameter of the Hamiltonian of the system, such as $H$, remains unsatisfactory\cite{Kapitulnik2019RMP, Wang2024RPP}.

In this work, we explore the emergence of SC in \stf\, by extending the temperature range of measurements by two orders of magnitude, down to $T=0.02$~K, and in perpendicular $H$ up to $18$~T.  Surprisingly, we find that the system fails to reach a zero-resistivity ($\rho=0$) state although SCFs are observed below the onset $T_{\mathrm{c}}$ within the insulator-metal phase coexistence region (Fig.~\ref{fig:Mixture}b).  Our results, however, reveal several features consistent with granular SC, reflecting the presence of superconducting domains embedded in the metallic percolating cluster (see sketch in Fig.~\ref{fig:Mixture}b).  Using percolation theory\cite{Kirkpatrick1973}, we extract the resistivity of the superconducting domains $\rho_{\mathrm{sc}}(T,H)$ and show that it exhibits quantum criticality associated with a field-tuned quantum superconductor-to-metal transition (QSMT).  At high $H$ where SC is suppressed, we observe universal conductance fluctuations (UCFs) in the millimeter-sized samples, consistent with a picture of mesoscopic phase segregation.  Our results thus reveal that failed SC in \stf\, arises due to the intrinsic inhomogeneity characteristic of the phase coexistence region, providing a new perspective on the ``anomalous metal'', i.e. failed superconductor behavior reported in diverse 2D superconducting systems\cite{Kapitulnik2019RMP, Wang2024RPP}.  
\\
\vspace{-12pt}

\noindent\textbf{Failed superconductivity}\\
The in-plane ($bc$) resistivity $\rho$ was measured on several single-crystalline \stf\, samples with a nominal $x=0.10-0.16$ (see Methods), encompassing the phase coexistence and FL regions (Fig.~\ref{fig:Mixture}b).  The maximum in $\rho(T)$ (Fig.~\ref{fig:Mixture}c and Supplementary Figs.~1b-d) identifies the Brinkman-Rice temperature $T_{\mathrm{BR}}$, above which metallic transport is lost and quasiparticles are destroyed\cite{Pustogow2021FL}.  FL behavior $\rho=\rho_0 + AT^2$ is observed at $T< T_{\mathrm{FL}}$ for higher $x$ (Fig.~\ref{fig:Mixture}c lower inset and Supplementary Fig.~1d).  Within the phase coexistence region, which is the focus of our work, $\rho(T)$ exhibits a drop with decreasing $T$ associated with the onset of SCFs at $T_{\mathrm{c}}$ (Fig.~\ref{fig:Mixture}c insets and Supplementary Fig.~1c), as discussed in detail below.  The values of $T_{\mathrm{BR}}$, $T_{\mathrm{FL}}$, and $T_{\mathrm{c}}$ are in good agreement with those from prior work\cite{Pustogow2021FL, Pustogow2021, Saito2021crystl} (see solid vs open symbols in Fig.~\ref{fig:Mixture}b).  

\begin{figure}[!tb]
\centering
\includegraphics[width=\columnwidth]{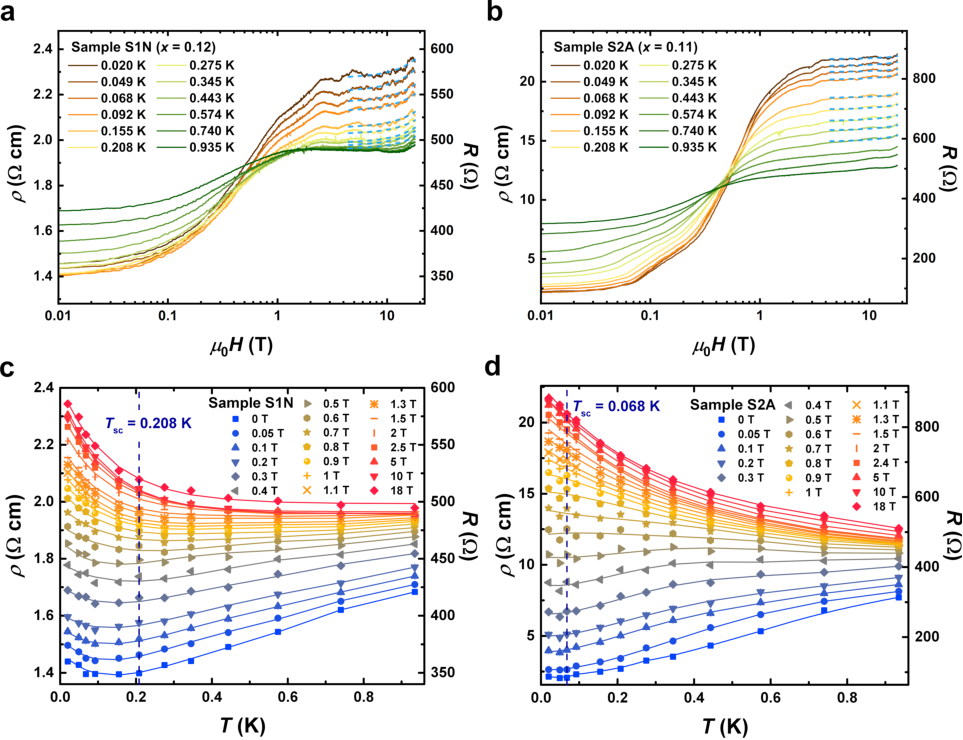}
\caption{\textbf{Failed superconductivity in the phase coexistence region.
} 
\textbf{a, b} $\rho$ (and resistance $R$) vs $H\perp bc$ for samples S1N and S2A, respectively, at several $T$, as shown.  The SCFs are suppressed for $H\gtrsim 2.5$~T.  Dashed lines are fits $\rho=\alpha + \beta H^2$ for 4~T$\leq H\leq 18$~T at $T<0.5$~K, representing the contributions from normal state transport.  \textbf{c, d} $\rho(T)$ extracted from the data in \textbf{a, b} for several $H$ up to 18~T, as shown.  The solid lines guide the eye. Despite the presence of SCFs, the zero-field $\rho(T)$ stops decreasing below $T_\mathrm{sc}$, and it either saturates (\textbf{d}) or shows an upturn (\textbf{c}).  This is referred to as ``failed superconductivity''.  $T_\mathrm{sc}=0.208$~K for sample S1N and $T_\mathrm{sc}=0.068$~K for sample S2A.
}
\label{fig:rawMR}
\end{figure}

Measurements down to the lowest $T$ show that, at $T<T_{\mathrm{c}}$, a large positive magnetoresistance (MR) $\rho(H)$ is observed in fields up to $H\sim 2.5$~T (Fig.~\ref{fig:rawMR}a,~b and Supplementary Fig.~2a,~b), indicating a suppression of SC by $H$.  At higher $H$, the MR is weak, representing the normal-state behavior.
%
%
The $\rho(T)$ curves extracted from the same data (Fig.~\ref{fig:rawMR}c,~d and Supplementary Fig.~2c,~d) show the $H$-dependence reminiscent of a field-tuned superconducting QPT.  However, despite the presence of SCFs, the global SC fails to set in: $\rho(T, H=0)$ does not drop all the way to zero but rather it saturates (Fig.~\ref{fig:rawMR}d) or shows an upturn (Fig.~\ref{fig:rawMR}c) at $T<T_\mathrm{sc}$, similar to the behavior of granular superconductors\cite{Kapitulnik2019RMP, CIQPT, Eley2011}.  The resistivity data thus suggest that SC is established at $T\approx T_\mathrm{sc}$ in superconducting islands (grains or domains) with negligible coupling to each other within the normal-state background, and it is the latter that determines $\rho(T=0, H=0)$.
\\
\vspace{-12pt}

\noindent\textbf{Revealing superconductivity: Percolation and quantum criticality}\\ 
Previous joint theoretical and experimental work has demonstrated the dominance of percolative effects in the phase coexistence region\cite{Pustogow2021}.  Here we focus on using classical percolation theory to extract the resistivity of the superconducting domains $\rho_{\mathrm{sc}}(T, H)$ from the measured $\rho(T, H)$.  For simplicity, we assume that the sample is a mixture of two types of domains: superconducting, with $\rho_{\mathrm{sc}}$, and normal-state, with resistivity $\rho_{\mathrm{n}}$.  According to the effective-medium approximation in 2D (ref.~\onlinecite{Kirkpatrick1973}), 
\begin{equation}
\frac{1/\rho_{\mathrm{sc}}-1/\rho}{1/\rho_{\mathrm{sc}}+1/\rho}V+\frac{1/\rho_\mathrm{n}-1/\rho}{1/\rho_\mathrm{n}+1/\rho}(1-V)=0,
\label{eqn:mixture}
\end{equation}
where $V$ is the superconducting domain volume fraction.  For clarity, we stress that here the superconducting domains are parts of the sample in which $\rho_{\mathrm{sc}}(T\leq T_{\mathrm{sc}}, H=0)=0$, i.e. $V$ is constant, while $\rho_{\mathrm{sc}}$ evolves with $H$ and $T$.  At $H\gtrsim 2.5$~T, where SCFs are suppressed, $\rho_{\mathrm{sc}}$ thus becomes the same as $\rho_{\mathrm{n}}$, i.e. $\rho_\mathrm{sc}\equiv\rho_\mathrm{n}\equiv\rho$ for $H\geq 2.5$~T.  Since $\rho_{\mathrm{n}}(H)$ is weak compared to the effect of $H$ on SCFs (Fig.~\ref{fig:rawMR}a,~b), the $H$-dependence of $\rho_\mathrm{n}$ can be neglected to study the low-field regime where SCFs are present.  Therefore, by assuming $\rho_\mathrm{n}(T, H< 2.5$~T$)=\rho(T, H=2.5$~T), from Eq.~\ref{eqn:mixture} at $T=T_\mathrm{sc}$ it follows that $V=1/2-[\rho(T = T_\mathrm{sc}, H=0)]/[2\rho(T = T_\mathrm{sc}, H=2.5~\mathrm{T})]$.  Finally, by rearranging Eq.~\ref{eqn:mixture}, we find $\rho_\mathrm{sc}(T, H)=\rho(T, H)[(2V-1)\rho_\mathrm{n}(T)+\rho(T, H)]/[(2V-1)\rho(T, H)+\rho_\mathrm{n}(T)]$.  Hence, for $H< 2.5$~T, both the volume fraction of the superconducting domains $V$ and their $\rho_\mathrm{sc}(T, H)$ can be calculated from the measured $\rho(T, H)$.

\begin{figure}[!tb]
\centering
\includegraphics[width=0.95\columnwidth]{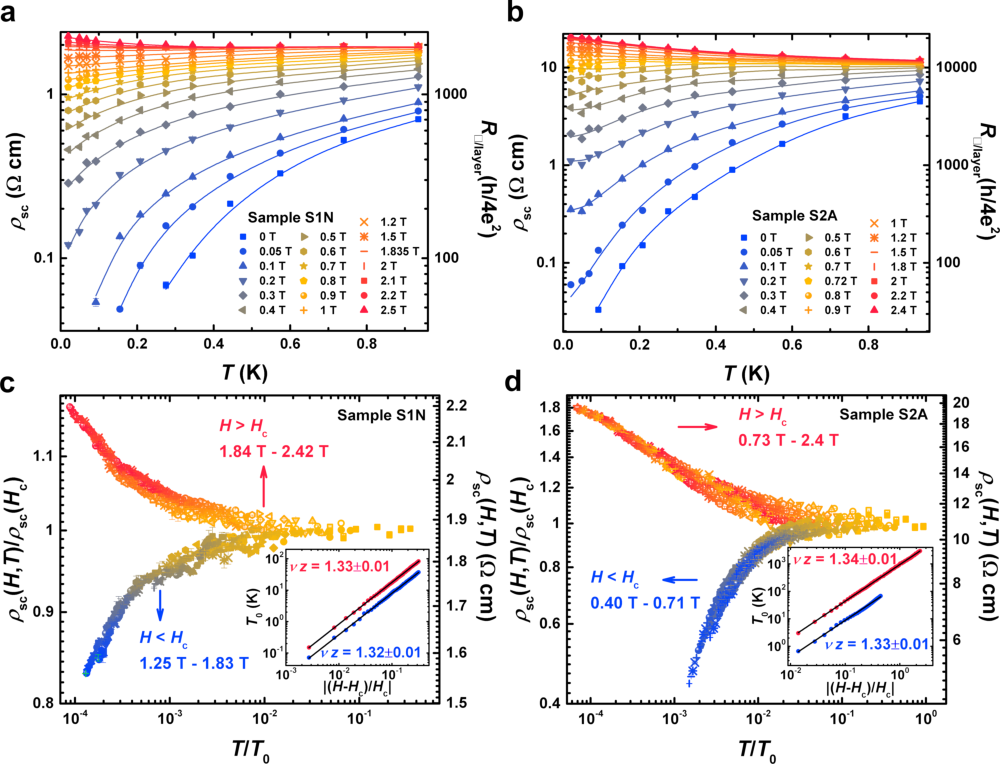}
\caption{\textbf{Resistivity of the superconducting domains and quantum criticality.} 
\textbf{a, b} $\rho_\mathrm{sc}(T)$ at several $H\leq 2.5$~T, as shown, for samples S1N ($x=0.12$) and S2A ($x=0.11$), respectively.  The data exhibit a change of the sign of $d\rho_\mathrm{sc}/dT$ at $H=H_{\mathrm{c}}$; $H_{\mathrm{c}}=1.835$~T for S1N, and $H_{\mathrm{c}}=0.720$~T for S2A.  Right axes: the corresponding sheet resistance per layer, $R_{\square/\mathrm{layer}}$, in units of quantum resistance for Cooper pairs, $R_\mathrm{Q}=h/4e^2$.  \textbf{c, d} Scaling of the $T\gtrsim 0.05$~K data (see also Methods) in \textbf{a} and \textbf{b}, respectively, with a single variable $T/T_0$.  The color and shape of the symbols correspond to the data for different $H$ near $H_\mathrm{c}$ in steps of 0.01~T.  Insets: The scaling parameter $T_0$ as a function of $|\delta|=|H-H_{\mathrm{c}}|/H_{\mathrm{c}}$ on both sides of $H_{\mathrm{c}}$.  For clarity, the two sets of $T_0$ are shifted vertically by an arbitrary amount.  The lines are linear fits with slopes $z\nu\simeq 1.3$, as shown; $T_0\propto |\delta|^{z\nu}$.  
}
\label{fig:Percolation&Scaling}
\end{figure}

For samples S1N with $x=0.12$ and S2A with $x=0.11$ (Fig.~\ref{fig:rawMR}), we find $V=0.16$ and $0.45$, respectively.  Both values are below the $V_{\mathrm{c}}=0.5$ percolation threshold in 2D (ref.~\onlinecite{Kirkpatrick1973}), indicating that superconducting domains are isolated from each other.  The $\rho_\mathrm{sc}(T)$ curves derived for $H\leq 2.5$~T are shown in Fig.~\ref{fig:Percolation&Scaling}a,~b (see also Supplementary Fig.~3a,~b).  
%
%
In contrast to $\rho(T, H=0)$ (Fig.~\ref{fig:rawMR}c, d), $\rho_\mathrm{sc}(T, H=0)$ exhibits an orders-of-magnitude drop with decreasing $T$, similar to other superconducting systems\cite{CIQPT}.  Furthermore, there is a field $H_\mathrm{c}$ where $\rho_\mathrm{sc}(T, H=H_\mathrm{c})$ is independent of $T$, suggesting\cite{Fisher1990} a field-tuned superconducting QPT in 2D.  Indeed, we find that, near $H_\mathrm{c}$, $\rho_\mathrm{sc}(T)$ for different $H$ can be collapsed onto a single function by rescaling the temperature, as shown in Fig.~\ref{fig:Percolation&Scaling}c,~d (see also Supplementary Fig.~3c,~d).  Therefore, $\rho_\mathrm{sc}(T, H) = \rho_{\mathrm{sc}}(H=H_\mathrm{c})f(T/T_0)$, where $T_0\propto |\delta|^{z\nu}$ holds over a wide, two orders of magnitude range of $|\delta|$ (insets in Fig.~\ref{fig:Percolation&Scaling}c,~d and Supplementary Fig.~3c,~d); $|\delta|=|H-H_{\mathrm{c}}|/H_{\mathrm{c}}$, with $z$ and $\nu$ the dynamical and correlation length exponents, respectively\cite{CIQPT}.  We find the same $z\nu\approx 1.3$ on both sides of the transition, as well as in all samples despite the differences in the values of their $H_\mathrm{c}$ (Table~\ref{Tab:rho_SC}).  
\begin{table}
\centering
 \caption{Scaling analysis of the superconducting domain resistivity $\rho_\mathrm{sc}(T, H)$
 }
 \vspace*{6pt}
 \begin{tabular}{|c| c c c|} 
 \hline
  Sample & $H_\mathrm{c}$ & $z\nu$ for $H<H_\mathrm{c}$  &$z\nu$ for $H>H_\mathrm{c}$  \\ [0.5ex] 
 \hline
  S1N&$1.835$~T&$1.32\pm0.01$&$1.33\pm0.01$\\
\hline
  S1&$1.790$~T&$1.31\pm0.01$&$1.26\pm0.01$\\
 \hline
 S2A&$0.720$~T&$1.33\pm0.01$&$1.34\pm0.01$\\
 \hline
 S2B&$0.945$~T&$1.33\pm0.01$&$1.30\pm0.01$\\ 
 \hline
 \end{tabular}

\label{Tab:rho_SC}
\end{table}
Remarkably, the same exponent $z\nu$ was found in thin films of conventional superconductors, in particular, in the well-studied field-tuned QSMT in 2D $a$-MoGe thin films\cite{Yazdani1995, Kapitulnik2019RMP}.  To verify the reliability of our scaling analysis even further, we check whether it can be applied directly to $\rho(T, H)$ at $T>T_{\mathrm{sc}}$, where $\rho(T)$ curves also exhibit a change of the sign of $d\rho/dT$ with increasing $H$ (Fig.~\ref{fig:rawMR}c, d).  We find that, although an approximate collapse of those data can be achieved (Supplementary Fig.~4), the resulting values of $z\nu$ (Supplementary Table 1) differ significantly from sample to sample or even on the two sides of $H_\mathrm{c}$ in a single sample, and $T_0(\delta)$ does not necessarily obey a power law, all in disagreement with general expectations for a QPT\cite{Sachdev-QPT}.  Therefore, the consistent results of our scaling analysis and reasonable exponents strongly support our procedure and assumptions.

While there is no reliable theory for QSMT critical exponents, experiments on a field-tuned superconducting QPT have been generally interpreted\cite{Kapitulnik2019RMP} in the context of a scaling theory for a $T=0$ superconductor-insulator transition (SIT) in 2D driven by quantum phase fluctuations\cite{Fisher1990}, according to which the critical resistivity equals $R_\mathrm{Q}=h/4e^2$, the quantum resistance for Cooper pairs.  The measured critical resistivities are usually scattered around $h/4e^2$ to within an order of magnitude\cite{Breznay2017PRB}.  Having this in mind, 
the orders-of-magnitude higher $\rho_\mathrm{sc}(H=H_\mathrm{c})$ found from our analysis ($\sim10^{3}h/4e^2$ and $\sim10^{4}h/4e^2$ in Fig.~\ref{fig:Percolation&Scaling}a,~b, respectively) may seem surprising at first.  We point out, though, that this is a consequence of the metal-insulator phase coexistence.  In particular, it is known\cite{Saito2021crystl, Pustogow2021} that \stf\, samples with $x=0.11-0.12$, which are the focus of this study, find themselves just above the percolation threshold for metallic conduction.  While the volume fraction of the metallic domains in the sample is just above 0.5, electric current is carried mostly by the infinite, percolating metallic cluster, which has an arbitrarily small volume fraction compared to the entire sample\cite{Kirkpatrick1973}; the majority of the metallic clusters or domains remain unconnected within an insulating background (see sketch in Fig.~\ref{fig:Mixture}b).  Therefore, although a metallic $T$-dependence ($d\rho/dT>0$; $H=0$) is observed at $T<T_{\mathrm{BR}}$, large values of $\rho$, above the Ioffe-Regel-Mott limit, are found in the normal state\cite{Saito2021crystl, Pustogow2021} (also, e.g., Fig.~\ref{fig:Mixture}c, Supplementary Figs.~1b-d).  The existence of the infinite metallic cluster that dominates transport also justifies our assumption that, when superconducting domains appear within the metallic regime at low enough $T$, the sample may be considered a mixture of only two types of domains: superconducting islands and the metallic background.  However, since $\rho_\mathrm{sc}$ has been calculated using $\rho$ of the entire sample rather than that of the infinite metallic cluster, this leads to the significantly enhanced absolute values of $\rho_\mathrm{sc}$, while the $H$- and $T$-dependences remain unaffected.   

Finally, we note that the volume fraction $V$ found from our percolation analysis actually represents the ratio of the superconducting domains to that of the infinite metallic cluster, and thus it may be expected to be a very small fraction of the volume of the entire sample.  To determine the superconducting volume fraction independently, we have performed the ac susceptibility measurements on other samples with $x\approx 0.11$ (Methods).  The estimated superconducting volume fraction is $\lesssim 0.2 \%$ (Supplementary Fig.~5), consistent with the above percolation scenario.  In particular, such a small number suggests that superconducting domains are located mainly, if not entirely, in the infinite metallic cluster (see sketch in Fig.~\ref{fig:Mixture}b), rather than distributed in isolated metallic domains.
\\
\vspace{-12pt}

\begin{figure}[!tb]
\centering
\includegraphics[width=\columnwidth]{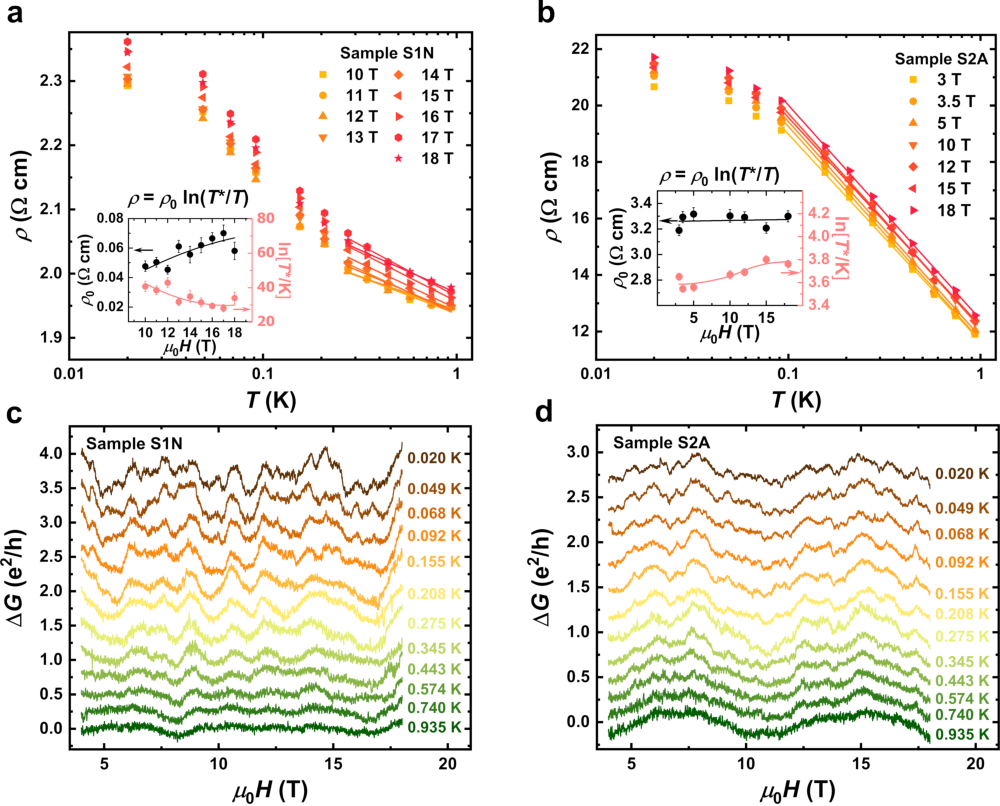}
\caption{\textbf{Properties of the field-revealed normal state in the phase coexistence region.} 
\textbf{a, b} The weak insulating behavior of the resistivity, $d\rho(T)/dT<0$, for samples S1N ($x=0.12$) and S2A ($x=0.11$), respectively, at high $H$ where SC is fully suppressed.  The solid lines are fits to $\rho=\rho_{0}(H) \ln(T^{\ast}(H)/T)$.  Insets: Fitting parameters $\rho_{0}(H)$ and $T^{\ast}(H)$.  The error bars correspond to $\pm 1$~SD of the fits.  \textbf{c, d} Conductance ($G=1/R$) fluctuations in samples S1N and S2A, respectively, from the data in Fig.~\ref{fig:rawMR}a,~b; $\Delta G(H)= [1/R(H)] - \langle G(H)\rangle$, where $\langle G(H)\rangle$ is the background obtained by third-order polynomial fitting for $4$~T~$\leq H\leq 18$~T (see Supplementary Note 1).  For clarity, $\Delta G(H)$ at different $T$ have been shifted vertically by an arbitrary amount.  For both samples, the root-mean-squared amplitude $\Delta G_{\mathrm{rms}}\sim 0.1 e^2/h$ at the lowest $T$.}
\label{fig:HighField}
\end{figure}

\noindent\textbf{Field-revealed normal state}\\
To determine the nature of the ground state revealed when SC is fully suppressed by $H\gtrsim 2.5$~T, we focus on $\rho(T)$ in the $T\rightarrow 0$ limit 
(Fig.~\ref{fig:HighField}a,~b), where positive MR is observed for both $x=0.11$ and $x=0.12$ (Fig.~\ref{fig:rawMR}a,~b).  A small, negative MR seen at 
%
%
higher $T$ ($T\gtrsim 0.5$~K) in intermediate $H$ up to $\sim9$~T for $x=0.12$ (Fig.~\ref{fig:rawMR}a) will be discussed elsewhere.  We find that, in the regime of positive MR, the data are described by a weak, insulating $\rho(T)=\rho_{0}\ln(T^{\ast}/T)$ (Fig.~\ref{fig:HighField}a,~b) over an order of magnitude in $T$ (Fig.~\ref{fig:HighField}b), followed by a stronger $T$-dependence at lower $T$ for $x=0.12$ (Fig.~\ref{fig:HighField}a).  At the lowest $T$, there is a tendency towards saturation  (Fig.~\ref{fig:HighField}a,~b), although at $T\lesssim 0.05$~K the effects of heating by the external noise cannot be completely ruled out  (see Methods).  Since $\rho(T)\propto\ln(1/T)$ is not suppressed by increasing $H$ and the MR is positive, both weak localization\cite{Lee1985} and the Kondo effect\cite{Kondo1964} can be ruled out as the origin of such weak insulating behavior.  In weakly disordered, homogeneous 2D metals, logarithmic corrections to conductivity and a positive MR can arise as a result of electron-electron interactions\cite{Lee1985}.  However, this is not consistent with our observation of a large ($\gtrsim 50$\% in Fig.~\ref{fig:HighField}b) $\ln(1/T)$ term in $\rho$, and plots of conductivity $\sigma=1/\rho$ vs $\ln T$ are not linear in the $T$ range over which $\rho(T)\propto\ln(1/T)$ holds (Supplementary Fig.~6).  A similar $\rho(T)\propto\ln(1/T)$ behavior was observed also in various granular materials\cite{Beloborodov2007} and in underdoped cuprates\cite{Ando1995PRL, Ono2000, Shi2020Sciadv}, but in both cases its precise microscopic origin is still not well understood.  In \stf, it is plausible that such $\rho(T)$ is indeed due to the intrinsic granularity of the metal-insulator phase coexistence region in the normal state.  The weak positive MR~$\propto H^2$ (see dashed lines in Fig.~\ref{fig:rawMR}a,~b) is thus attributed to the classical orbital effect in the metallic domains which dominate transport, i.e. to the same mechanism  that leads to the positive MR~$\propto H^2$  in other conventional, FL metals\cite{Pippard1989}.  

We also find that, in the normal state as $T\rightarrow 0$, the Hall resistivity is immeasurably small, i.e.,  $\rho_\mathrm{yx}(H)\approx 0$ and $\rho\equiv\rho_\mathrm{xx}\gg\rho_\mathrm{yx}$ (Supplementary Fig.~7).  While there are no prior Hall effect measurements on chemically substituted $\kappa$-organics, a study of insulating parent compounds at high $T\gtrsim 50$~K has shown\cite{Culo2019} a large, orders-of-magnitude decrease of the positive (holelike) Hall coefficient $R_\mathrm{H}= \rho_{\mathrm{yx}}(H)/H$ as the Mott MIT is approached by decreasing the effective correlation strength $U/W$.  Therefore, as $U/W$ is reduced even further with $x$ (i.e., with increasing $W$) and the Mott gap closes, the Hall coefficient may become immeasurably small.  However, a detailed study of the evolution of $R_\mathrm{H}$ with $x$ and temperature down to $T<1$~K is beyond the scope of this work.

It is intriguing that $\rho(H)$ in the field-revealed normal state (Fig.~\ref{fig:rawMR}a,~b) exhibits fluctuations that are highly reproducible upon sweeping $H$ up and down over the entire measurement range (Supplementary Fig.~8a).  Figure~\ref{fig:HighField}c,~d shows the corresponding conductance ($G=1/R$) fluctuations, $\Delta G(H)$, which are random, i.e. uncorrelated beyond a small field scale $H_\mathrm{1/2}\lesssim 0.5$~T (Supplementary Fig.~8b and Supplementary Note 1), and get more pronounced with decreasing $T$.   The phenomenology of these sample-specific magnetofingerprints is consistent with UCFs\cite{LeeStone1985, LeeStone1987PRB}, which are a consequence of quantum interference of electron waves propagating along different paths, a manifestation of quantum coherent transport.  In an applied $H$, the flux modifies the relative phases of interfering paths via the Aharonov-Bohm effect, leading to reproducible conductance fluctuations as a function of $H$.  Such fluctuations in metallic samples have a universal $T=0$ root-mean-squared amplitude $\Delta G_\mathrm{rms}\sim e^2/h$ in all dimensions, and independent of sample size and degree of (weak) disorder.  However, when the phase coherence length (i.e. inelastic scattering length $L_\mathrm{in}$) becomes smaller than the sample size, $\Delta G_\mathrm{rms}$ decreases with increasing $T$ due to classical self-averaging at scales larger than $L_\mathrm{in}$.  In addition, $\Delta G_\mathrm{rms}$ can be reduced below its $T=0$ value due to thermal averaging, i.e. because charge carriers with a different momentum acquire different phases along the same path.  For \stf, we estimate (Supplementary Note 1) that phase is thermally randomized over a length $L_\mathrm{T}\sim 1000$~nm at the lowest $T$.  On the other hand, $L_\mathrm{in}\sim 100$~nm~$\ll L_\mathrm{T}$, and thus thermal averaging can be neglected.  We find that $\Delta G_{\mathrm{rms}}\sim 0.1 e^2/h$ observed at the lowest $T$ (Fig.~\ref{fig:HighField}c,~d) is indeed consistent with classical self-averaging (see Supplementary Note 1 for a more detailed discussion).  The surprising observation of UCFs in macroscopic, millimeter-sized samples is thus attributed to transport through the metallic, FL regions that surround insulating domains, i.e., it is consistent with the picture of phase coexistence just above the percolation threshold for metallic conduction.  However, the precise theoretical description of UCFs in inhomogeneous systems, especially in the presence of Mott droplets or domains, remains an open question.
\\
\vspace{-12pt}

\noindent\textbf{Discussion}\\
\noindent  Our findings in highly frustrated \stf\, paint a consistent picture of an intrinsically granular superconductor, which we expect within a phase coexistence region.  However, in contrast to the behavior of pressure-tuned parent compounds in which a superconducting phase arises on the metallic side\cite{Powell2011, dressel2020advances}, here SC emerges only within a phase coexistence region and it fails to develop global phase coherence even as $T\rightarrow 0$.  We attribute the fragility of the superconducting state in \stf\, to disorder that is enhanced by chemical substitution\cite{Saito2018}.  By taking into account percolative effects, already known to be present in the phase coexistence region\cite{Pustogow2021, dressel2020advances}, we have demonstrated that the observed failed superconductivity is a result of the phase coexistence of a superconductor, the FL metal, and the Mott insulator, such that isolated superconducting domains are concentrated in the infinite metallic cluster that barely percolates in the insulating background (see sketch in Fig.~\ref{fig:Mixture}b).  It is interesting that a percolative scenario of the interplay of superconductivity and the intrinsic inhomogeneity has been argued for cuprates\cite{Pelc2018, Pelc2019}, and suggested\cite{Saito2021crystl} also for fullerides and other unconventional superconductors exhibiting a dome of $T_\mathrm{c}$.

The percolation analysis has allowed us also to reveal evidence for quantum-critical behavior of $\rho_\mathrm{sc}(T, H)$, the resistivity of superconducting islands, reflecting an underlying $H$-tuned QSMT with the critical exponent $z\nu\approx 1.3$.  Assuming that $z=1$ due to the long-range Coulomb interaction between charges\cite{Fisher1990, FisherGG1990}, as measured, e.g., in 2D $a$-MoGe thin films\cite{Yazdani1995}, we find that $\nu\approx1.3$.  While there is currently no satisfactory theory for a field-tuned QSMT\cite{Kapitulnik2019RMP}, at least in the context of the 2D SIT the exponent $\nu\geqslant1$ is believed to be in the universality class of the $T=0$ SIT in a disordered system\cite{Fisher1990}, consistent with our conclusions about the importance of disorder in \stf.  Similar scaling behavior with $z\nu> 1$ was found also in underdoped La-214 cuprates\cite{Shi2014NatPhy, Shi2020NatCom}, and attributed to the interplay of disorder and quantum phase fluctuations.

At low fields $H<H_\mathrm{c}$ and low temperatures where superconducting correlations are present, the observed $\rho_\mathrm{xx}\neq 0$ with $\rho_\mathrm{yx}(H)\approx 0$ (Supplementary Fig.~7) is reminiscent of the behavior found within the vortex liquid regime of some conventional disordered 2D superconductors\cite{Kapitulnik2019RMP, Breznay2017SciAdv} and oxide interfaces\cite{Chen2021}, as well as in cuprates, including YBa$_2$Cu$_3$O$_y$ thin films\cite{Yang2019}, charge- and spin-stripe ordered La$_{1.875}$Ba$_{0.125}$CuO$_4$ (ref.~\cite{Li2019}), La$_{1.7}$Eu$_{0.2}$Sr$_{0.1}$CuO$_4$ and La$_{1.48}$Nd$_{0.4}$Sr$_{0.12}$CuO$_4$ (ref.~\cite{Shi2021}), and Bi-2201 from underdoped to heavily overdoped regions\cite{Terzic2025}.  While $\rho_\mathrm{yx}(H)\approx 0$ accompanied by $\rho_\mathrm{xx}(T\rightarrow 0)= 0$ has been determined\cite{Shi2014NatPhy, Shi2020Sciadv, Shi2020NatCom, Shi2021, Terzic2025} to be a consequence of the slowing down and freezing of the vortex motion with decreasing $T$ in the presence of disorder, the origin of $\rho_\mathrm{yx}(H)\approx 0$ with $\rho_\mathrm{xx}(T\rightarrow 0)\neq 0$, a characteristic of ``anomalous metals'' or ``failed superconductors'', is still an open question\cite{Kapitulnik2019RMP, Delacretaz2018}.  In \stf, we have explained $\rho_\mathrm{xx}(T\rightarrow 0)\neq 0$ (Fig.~\ref{fig:rawMR}c,~d) by percolative effects, i.e., the presence of isolated superconducting islands, so that the measured $\rho_\mathrm{yx}(H)$ must then correspond to the Hall resistivity of the normal background; hence, $\rho_\mathrm{yx}(H)\approx 0$.  Therefore, the remaining question is the origin of the vanishing Hall response in the normal state over a wide range of $H$.  Interestingly, $\rho_\mathrm{yx}(H)\approx 0$ was observed\cite{Shi2021} also in the field-revealed normal state of La$_{1.7}$Eu$_{0.2}$Sr$_{0.1}$CuO$_4$ and La$_{1.48}$Nd$_{0.4}$Sr$_{0.12}$CuO$_4$ over a huge range of $H$ and $T$, where it seemed to be crucially related to the presence of short-range charge-order domains or stripes, separated by hole-poor spin stripes.  The similarity to our simpler system,  the metal-insulator phase coexistence region in \stf, suggests that $\rho_\mathrm{yx}(H)\approx 0$ might be a more general property of systems with an intrinsic electronic inhomogeneity.  Further insight into this issue might come from experiments on other materials and from other techniques, especially local probes such as STM.
\\
\vspace*{-12pt}

\begin{methods}
\vspace{-12pt}
\noindent\textbf{Samples.}  Plate-like single crystals of \STF\, with various stoichiometries were prepared by electrochemical oxidation\cite{Geiser1991}; here (BEDT-TTF) and (BEDT-STF) stand for bis-(ethylenedithio)-tetrathiafulvalene and bis-(ethylenedithio)-diseleniumdithiafulvalene, respectively.   Measurements of transport in the $bc$ plane, with current along the $c$-axis (the most conductive axis\cite{Iniguez2014PRB}), were performed on crystals with $x=0.10 - 0.16$ at $1.5$~K$\leq T \leq 300$~K (Fig.~\ref{fig:Mixture}b,~c and Supplementary Fig.~1b-d).  Detailed magnetotransport measurements down to $T=0.02$~K were carried out on samples with $x=0.11\pm0.01$ and $x=0.12\pm0.01$.  After $\sim2$~years, the absolute value of the resistance of sample S1 ($x=0.12$) changed, so we consider it a different sample S1N ($x=0.12$).  The dimensions of the S1 (S1N) sample are 
$a \times b \times c = 0.04 \times 0.4 \times 0.9~\textrm{mm}^3$.  The data are shown for the voltage contacts (``V+'' and ``V--'' in the sample image in Suppl. Fig.~1a) separated by $\approx0.44~\textrm{mm}$. The dimensions of sample S2 ($x=0.11$) are $a \times b \times c=0.13 \times 0.7 \times 1.3 ~\textrm{mm}^3$, with a pair of voltage contacts on each side of the sample, separated by $\approx0.37$~mm and $\approx0.36$~mm, respectively.  The resistivity data obtained from the two pairs are qualitatively, and nearly quantitatively consistent with each other; nevertheless, for precision, we denote them as samples S2A and S2B.  All contacts are made using a DOTITE XC-12 carbon paste connected to $25 ~\mu\textrm{m}$-thick gold wires. \\
\vspace{-12pt}

\noindent 
The ac susceptibility measurements were carried out on dozens of \stf\, samples with a nominal $x=0.11$ and dimensions comparable to those used for transport.  The total volume of these samples is about $0.63~\textrm{mm}^3$.  To calibrate the ac susceptibility measurements, amorphous indium pieces with twice the total volume ($1.26~\textrm{mm}^3$) and similar shapes to the \stf\, samples were prepared.  \\
\vspace*{-12pt}

\noindent\textbf{Measurements.}  The longitudinal resistance and Hall effect were measured using the standard four-probe ac method (typically $\sim13$~Hz) with the Lakeshore 372 AC resistance bridge or Signal Recovery 7265 lock-in amplifier and SR5113 voltage preamplifier.  Depending on the temperature and sample, the excitation current (density) ranged from $10~\textrm{nA}$ ($\sim10^{-5}$Acm$^{-2}$) at the lowest $T$, to 100~nA--1~$\mu$A (up to $\sim10^{-3}$~Acm$^{-2}$) at the highest $T$.  These excitation currents were low enough to avoid Joule heating.  Magnetic field was applied parallel and antiparallel to the $a$ axis, and  the Hall resistance was determined from the transverse voltage by extracting the component antisymmetric in the magnetic field.  To increase the signal-to-noise ratio, an excitation of $10~\mu\textrm{A}$ ($\sim10^{-2}$~Acm$^{-2}$) was used in the Hall effect measurements.  Despite potential heating, this does not affect the main conclusion that the Hall resistivity $\rho_{\mathrm{yx}}\approx 0$ and, in particular, it is negligible compared to the longitudinal resistivity.  The sheet resistance per layer $R_{\square/\mathrm{layer}}=\rho/d$, where $d\sim 16$~\AA\, is the interlayer distance\cite{Saito2021JMCC}.\\ 
\vspace*{-12pt}

\noindent 
Several different cryostats were used for the resistivity measurements, including a $^3$He system ($0.25~\textrm{K} \le T \le 300 ~\textrm{K}$) and a dilution refrigerator ($0.02~\textrm{K} \le T \le 1 ~\textrm{K}$) equipped with superconducting magnets up to 9~T and 18~T, respectively.  A 1~k$\Omega$ resistor in series with a commercial $\pi$ filter [Newark 19F687 with 5~dB (60~dB) noise suppression at 10~MHz (1~GHz)] was placed in each wire at the room temperature end of these cryostats to reduce the noise and heating by EM radiation.  It was verified that adding another $\pi$ filter [35 dB (70 dB) noise suppression at 10 MHz ($>100$~MHz)] in series did not affect the results down to the lowest test $T=0.070$~K, indicating that the effects of heating by the external EM noise are negligible at least down to that temperature.  At the same time, they cannot be completely ruled out at even lower $T$ and, indeed, it is likely because of the heating that most of the data at $T\lesssim 0.05$~K could not be collapsed in Fig.~\ref{fig:Percolation&Scaling}c,~d and Supplementary Fig.~3c, d.  Other measurements were performed in a variable-temperature insert ($1.5 ~\textrm{K} \le T \le 300 ~\textrm{K}$).  The Hall effect was measured in a dilution refrigerator ($0.05 ~\textrm{K} \le T \le 1 ~\textrm{K}$) with a 9~T superconducting magnet and low-pass $RC$ filters with $R=1$~k$\Omega$ and $C=10$~nF installed on each wire at the mixing chamber stage of the dilution refrigerator.  In all cryostats, the fields were swept at constant temperatures using $0.1-0.3~\textrm{T/min}$ sweep rates that were low enough to avoid eddy current heating of the samples.  Zero-field $\rho(T)$ curves shown in Fig.~\ref{fig:Mixture}c and Supplementary Figs.~1b-d were obtained by sweeping temperature at $<2~\textrm{K/min}$.  In all other measurements, the $T$-dependence was obtained by waiting at each given bath $T$ long enough for the sample temperature to equilibrate; the wait times were $\sim4-20$~minutes depending on the temperature and the cryostat.
\\
\vspace*{-12pt}

\noindent The ac susceptibility was measured in a $^3$He system ($0.3~\textrm{K} \le T \le 60 ~\textrm{K}$) using the conventional mutual inductance technique with a home-made ac susceptometer.  An ac magnetic field ($1.6~\textrm{Oe}$, $200~\textrm{Hz}$) was generated by an ac current source (Stanford Research CS 580), while a lock-in amplifier (Stanford Research SR 860) was used to measure the corresponding signal from a balanced pick-up coil.  Dozens of \stf\, samples with $x=0.11$ and a total volume $\sim1.26~\textrm{mm}^3$ were placed inside the pick-coil with random orientations with respect to the ac magnetic field. Consistent results were obtained by cooling down ($\le0.5~\textrm{K/min}$ from $\sim5~\textrm{K}$ to $\sim0.3~\textrm{K}$) and warming up.  The ac susceptometer was calibrated by measuring the superconducting transition of indium that was cut to have twice the volume 
and similar shape as the \stf\, samples.  
\\

\end{methods}

\vspace*{-12pt}

\noindent{\Large{\textbf{Data availability}}}  

\noindent The data that support the findings of this study are available within the paper and the Supplementary Information.  Additional data related to this work are available from the corresponding author upon request.

\bibliography{literature}

\bibliography{scibib}

\bibliographystyle{Science}

\vspace*{6pt}

\noindent{\Large{\textbf{Acknowledgements}}}

\noindent We thank G. Untereiner for assistance with sample preparation.  This work was supported by NSF Grants Nos. DMR-2104193 (D. P.),  DMR-1707785 (D. P.), DMR-2409911 (V. D.), and the National High Magnetic Field Laboratory through the NSF Cooperative Agreement Nos. DMR-2128556 and NSF DMR-1644779, and the State of Florida.  \\
\vspace*{-6pt}

\noindent{\Large{\textbf{Author contributions}}}

\noindent Single crystals were grown and prepared by Y.~S. and A.~K.;
Y.~W. and E.~S.~C. performed the measurements and analyzed the data; V.~D. contributed to the analysis and interpretation of the data; Y.~W., V.~D., A.~P., M.~D., and D.~P. discussed the results and the interpretation; Y.~W., V.~D., and D.~P. wrote the manuscript, with input from all authors; D.~P. planned and supervised the investigation.\\
\vspace*{-6pt}

\noindent{\Large{\textbf{Competing financial interests}}}

\noindent The authors declare no competing interests.\\
\vspace*{-6pt}

\noindent{\Large{\textbf{Additional information}}}

\noindent\textbf{Supplementary information} accompanies this paper.  

\noindent\textbf{Correspondence} and requests for materials should be addressed to D.P.~(email: dragana@magnet.fsu.edu).


\clearpage

\noindent\textbf{\Large{Supplementary Information}}
\vspace*{12pt}

\renewcommand{\figurename}{{\bf{Supplementary Fig.}}}
\renewcommand{\tablename}{{\bf{Supplementary Table}}}

\makeatletter
\makeatletter \renewcommand{\fnum@figure}{{\bf{\figurename~\thefigure}}}
\makeatletter \renewcommand{\fnum@table}{{\bf{\tablename~\thetable}}}

\makeatother

\setcounter{figure}{0}
\setcounter{table}{0}

\baselineskip=24pt

\noindent{\large{\textbf{Supplementary Figures}}}\\
\vspace{12pt}

%
\begin{figure}[!h]
\centerline{\includegraphics[width=\textwidth]{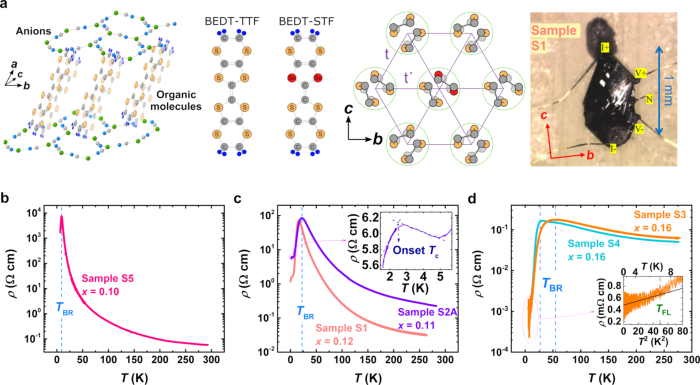}} 
\caption{\textbf{ $\bm{\kappa}$-STF$\bm{_x}$  samples.}  \textbf{a}    The crystal structure contains dimers of the organic donor molecules (BEDT-TTF) and (BEDT-STF) forming layers in the $bc$-plane, which are separated by the Cu$_2$(CN)$_3$ anion sheets.  In (BEDT-STF), two S atoms of the inner rings are replaced by Se.  One electron per dimer is donated to the anion sheets.  Each dimer carries one hole with spin $1/2$, such that within the $bc$ plane the dimers form a nearly ideal triangular lattice, with the interdimer transfer integrals $t'/t=0.83$ close to complete frustration.  The photo shows sample S1, with the $c$ axis along the direction from one current contact (``I+'' ) to another ( ``I--'').  \textbf{b-d} $\rho(T)$ for samples S5 ($x=0.10$), S1 ($x=0.12$) and S2A ($x=0.11$), S3 ($x=0.16$) and S4 ($x=0.16$), respectively.  The Brinkman-Rice temperatures $T_{\mathrm{BR}}$, marked by the vertical dashed lines, are determined from the peaks in $\rho(T)$.  ($T_{\mathrm{BR}}$ for sample S1 is shown in Fig.~1c.)  The inset in \textbf{c} shows the determination of $T_{\mathrm{c}}=(2.5\pm 0.3)$~K, the onset of superconducting fluctuations in sample S2A ($x=0.11$).  For this $x$, the Fermi-liquid behavior is not observed.  The inset in \textbf{d} shows the determination of  $T_{\mathrm{FL}}$ in sample S3.  For $x=0.16$, superconducting fluctuations are not observed down to the lowest measurement $T=0.270$~K.}
\label{fig:Cool-down}
\end{figure}

\clearpage

\begin{figure}[!t]
\centerline{\includegraphics[width=\textwidth]{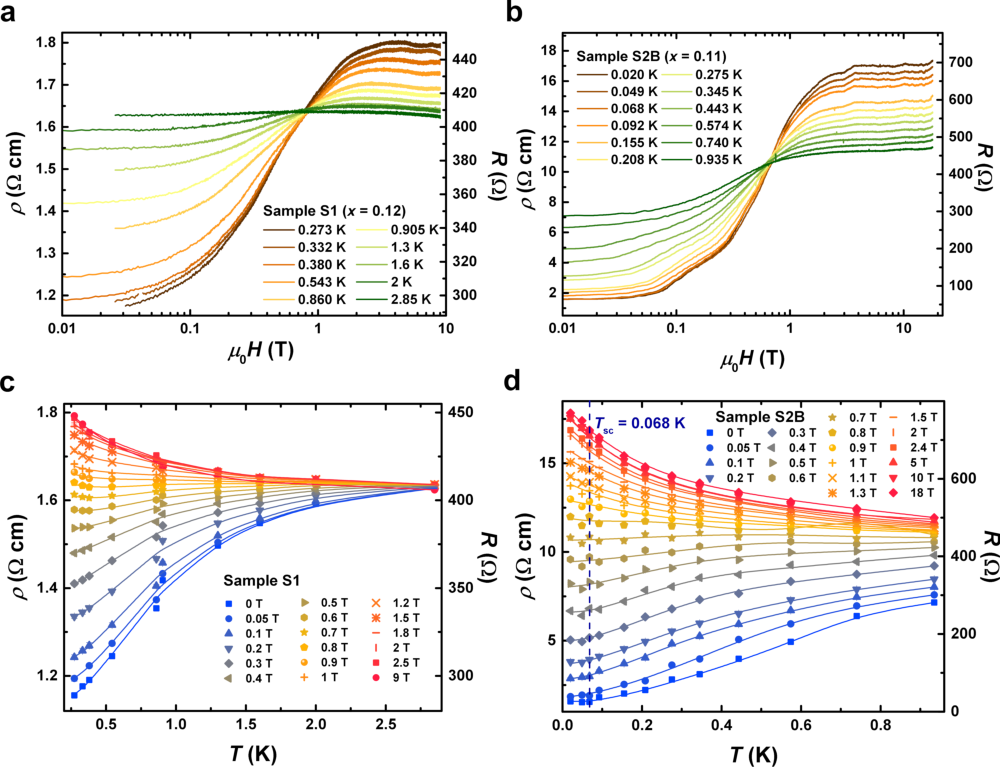}} 
\caption{\textbf{Failed superconductivity in the phase coexistence region.} 
$\rho$ vs $H\perp bc$ for samples S1 ($x=0.12$) and S2B ($x=0.11$), respectively, at several $T$, as shown.  The SCFs are suppressed for $H\gtrsim 2.5$~T.   \textbf{c, d} $\rho(T)$ extracted from the data in \textbf{a, b} for several $H$, as shown.  From both \textbf{a} and \textbf{c}, it is clear that the effect of $H$ becomes observable below the onset $T_{\mathrm{c}}\sim2.7$~K, indicating the suppression of SCFs at higher $T$.  Since sample S1 was measured only down to $0.273$~K, here no low-$T$ saturation or upturn of $\rho(T)$ is observed at $H=0$.   In \textbf{d}, $\rho(T, H=0)$ stops decreasing with decreasing $T$ below $T_\mathrm{sc}=0.068$~K.
}
\label{fig:raw_MR_supp}
\end{figure}

%

\clearpage
%
\begin{figure}[!t]
\centerline{\includegraphics[width=\textwidth]{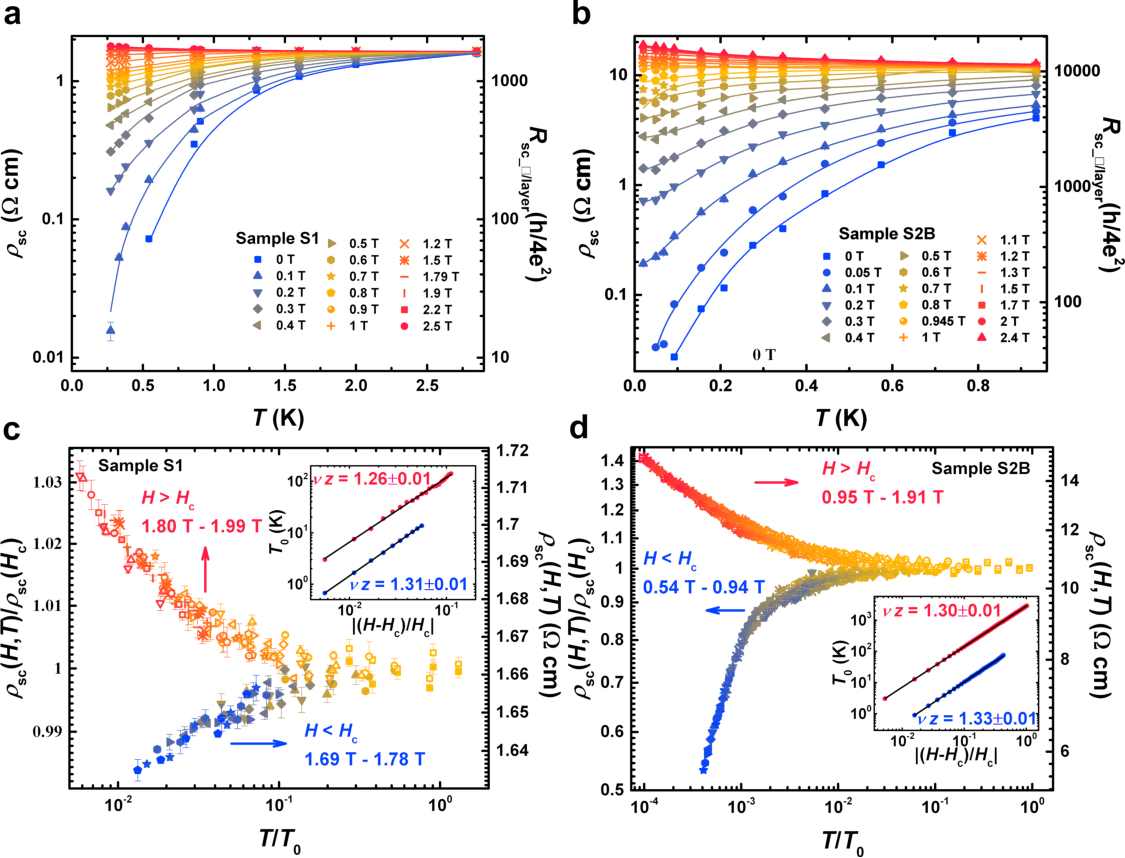}}
\caption{\textbf{Resistivity of the superconducting domains and quantum criticality.}  \textbf{a, b} $\rho_\mathrm{sc}(T)$ at several $H\leq 2.5$~T, as shown, for samples S1 and S2B, respectively.  Since sample S1 was measured only down to $0.273$~K and no low-$T$ saturation or upturn of $\rho(T)$ was observed at $H=0$ (Supplementary Fig.~\ref{fig:raw_MR_supp}c), its superconducting volume fraction is assumed to be the same as for sample S1N (see Methods), i.e. $V=0.16$.  For sample S2B, the percolation analysis gives $V=0.45$.  \textbf{c, d} Scaling of the $T\gtrsim 0.05$~K data (see also Methods) in \textbf{a} and \textbf{b}, respectively.  The color and shape of the symbols correspond to the data for different $H$ near $H_\mathrm{c}$ in steps of 0.01~T; $H_{\mathrm{c}}=1.790$~T for sample S1 and $H_{\mathrm{c}}=0.945$~T for sample S2B.  Insets: The scaling parameter $T_0$ as a function of $|\delta|=|H-H_{\mathrm{c}}|/H_{\mathrm{c}}$ on both sides of $H_{\mathrm{c}}$.  For clarity, the two sets of $T_0$ are shifted vertically by an arbitrary amount.  The lines are linear fits with slopes $z\nu\simeq 1.3$, as shown; $T_0\propto |\delta|^{z\nu}$.   
}
\label{fig:rho_sc_scaling_supp}
\end{figure}
%

\clearpage

\begin{figure}[!t]
\centerline{\includegraphics[width=0.85\textwidth]{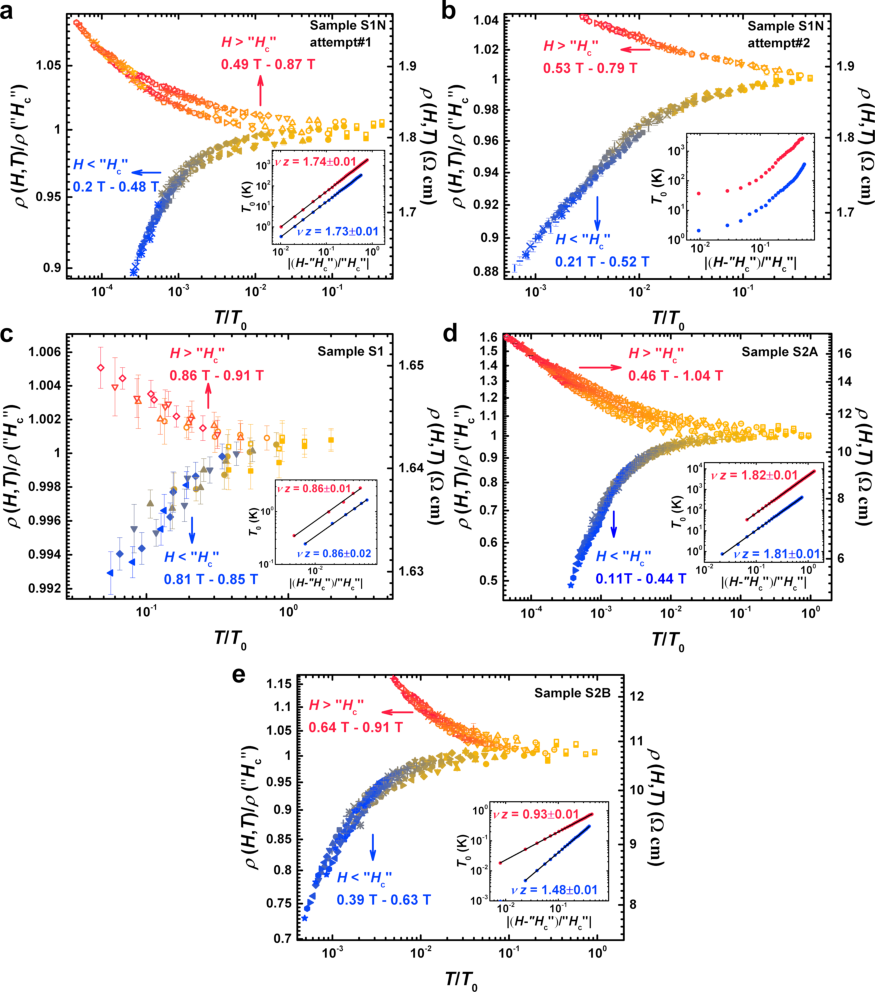}}
\caption{\textbf{Scaling analysis of the sample resistivity $\bm{\rho(T, H)}$.} 
\textbf{a, b} Two attempts to scale the data from Fig.~2c at $T>T_{\mathrm{sc}}$ with respect to a single variable $T/T_0$ using the nominal critical fields ``$H_\mathrm{c}$''$=0.485$~T and ``$H_\mathrm{c}$''$=0.525$~T, respectively.  \textbf{c, d, e} Scaling of the data from Supplementary Fig.~\ref{fig:raw_MR_supp}c, Fig.~2d (at $T>T_{\mathrm{sc}}$), and Supplementary Fig.~\ref{fig:raw_MR_supp}d (at $T>T_{\mathrm{sc}}$), respectively.  For all panels, the values of ``$H_c$'' are estimated as the fields where $\rho$ shows almost no $T$-dependence.  Insets: The scaling parameter $T_0$ as a function of $|\delta|=|H-$``$H_{\mathrm{c}}$''$|/$``$H_{\mathrm{c}}$'' on both sides of ``$H_{\mathrm{c}}$''.  For clarity, the two sets of $T_0$ are shifted vertically by an arbitrary amount.  The lines are linear fits with slopes $z\nu$, as shown.  Although an approximate collapse of the data can be achieved, it gives inconsistent results for $T_0(\delta)$, in disagreement with general expectations for a QPT.}
\label{fig:rho_scaling_supp}
\end{figure}
%

%
\clearpage

\begin{figure}[!h]
\centerline{\includegraphics[width=\textwidth]{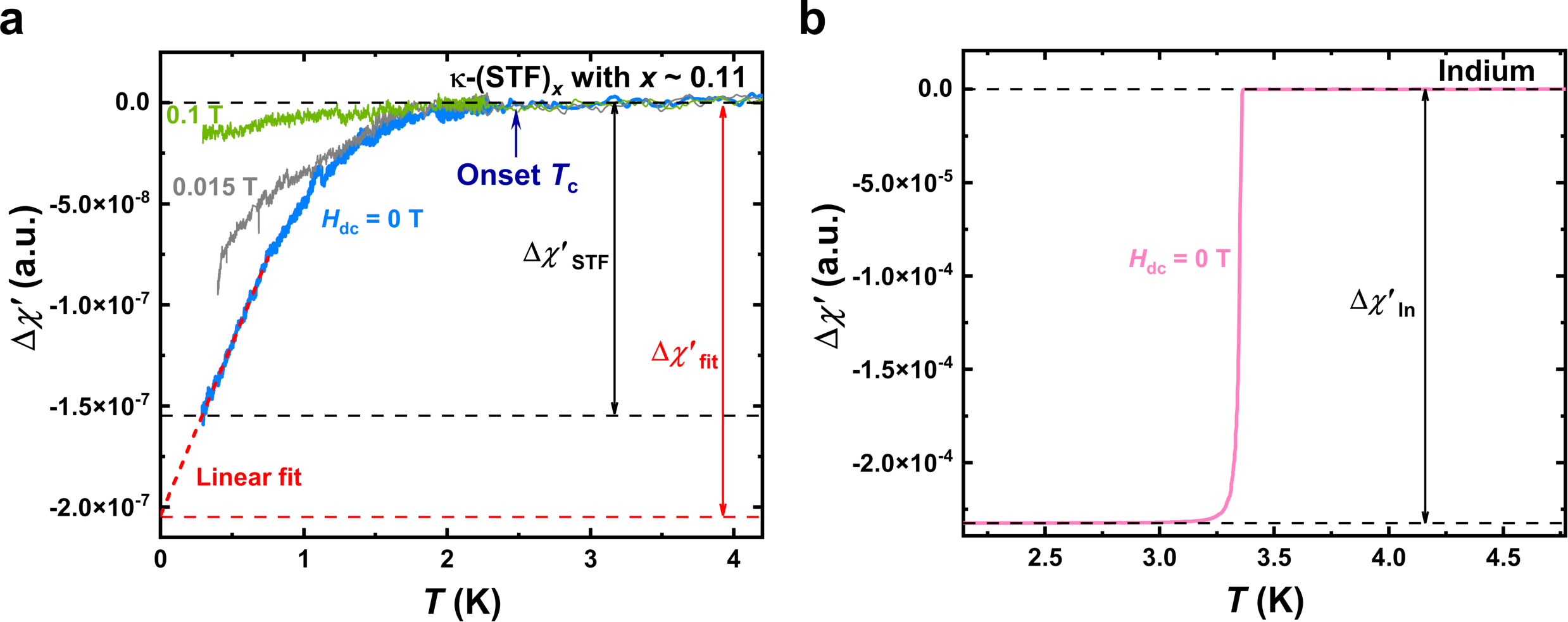}}
\caption{\textbf{Temperature-dependent ac susceptibility measurements.} 
\textbf{a} The change of the real component of the ac susceptibility of the \stf\, samples with nominal $x= 0.11$ measured by cooling down in dc fields $H_\mathrm{dc}= 0$~T, $0.015$~T, and $0.1$~T, as shown.  $\chi'_{\mathrm{STF}}(T)$ starts to drop at $T_{\mathrm{c}}=(2.5\pm 0.3)$~K, consistent with earlier magnetic susceptibility studies [S1] 
carried out down to 1.4~K. The magnitude of the drop is heavily reduced by $H_\mathrm{dc}$ up to $0.1~\mathrm{T}$, indicating, in fact, that the first critical field of the superconducting state is lower than 0.015~T.  However, since $\chi'_{\mathrm{STF}}(T)$ does not saturate at the lowest measurement $T\approx 0.3$~K, a linear fit (red dashed line) is performed to estimate the upper bound of the signal change in $H_\mathrm{dc}= 0$.  As shown in the figure, the change of the real component of the ac susceptibility for $H_\mathrm{dc}= 0$ is estimated to be $1.55\times 10^{-7}$--~$2.04\times 10^{-7}$ (a.u.).  \textbf{b} For calibration, the change of the real component of the ac susceptibility of the amorphous indium pieces was also measured by cooling down in $H_\mathrm{dc}= 0$. They were cut to have similar shapes as the \stf\, samples and with twice the volume.  The abrupt change $\Delta\chi'_{\mathrm{In}}\sim 2.32\times 10^{-4}~\textrm{a.u.}$  near $3.4$~K corresponds to the superconducting transition of indium. Since both \stf\, and indium were measured using the same pick-up coil, the ratio of their signal changes characterizes the ratio of their superconducting volumes. As indium is considered a pure superconductor, the superconducting volume fraction in \stf\, is estimated to be $0.12\%-0.18\%$.  Although measuring the superconducting volume fraction from the magnetic susceptibility is known to be not precise [S2], 
it should give a good order-of-magnitude estimate.}
\label{fig:ac-susc}
\end{figure}
%

%
\begin{figure}[!tb]
\centerline{\includegraphics[width=0.53\textwidth]{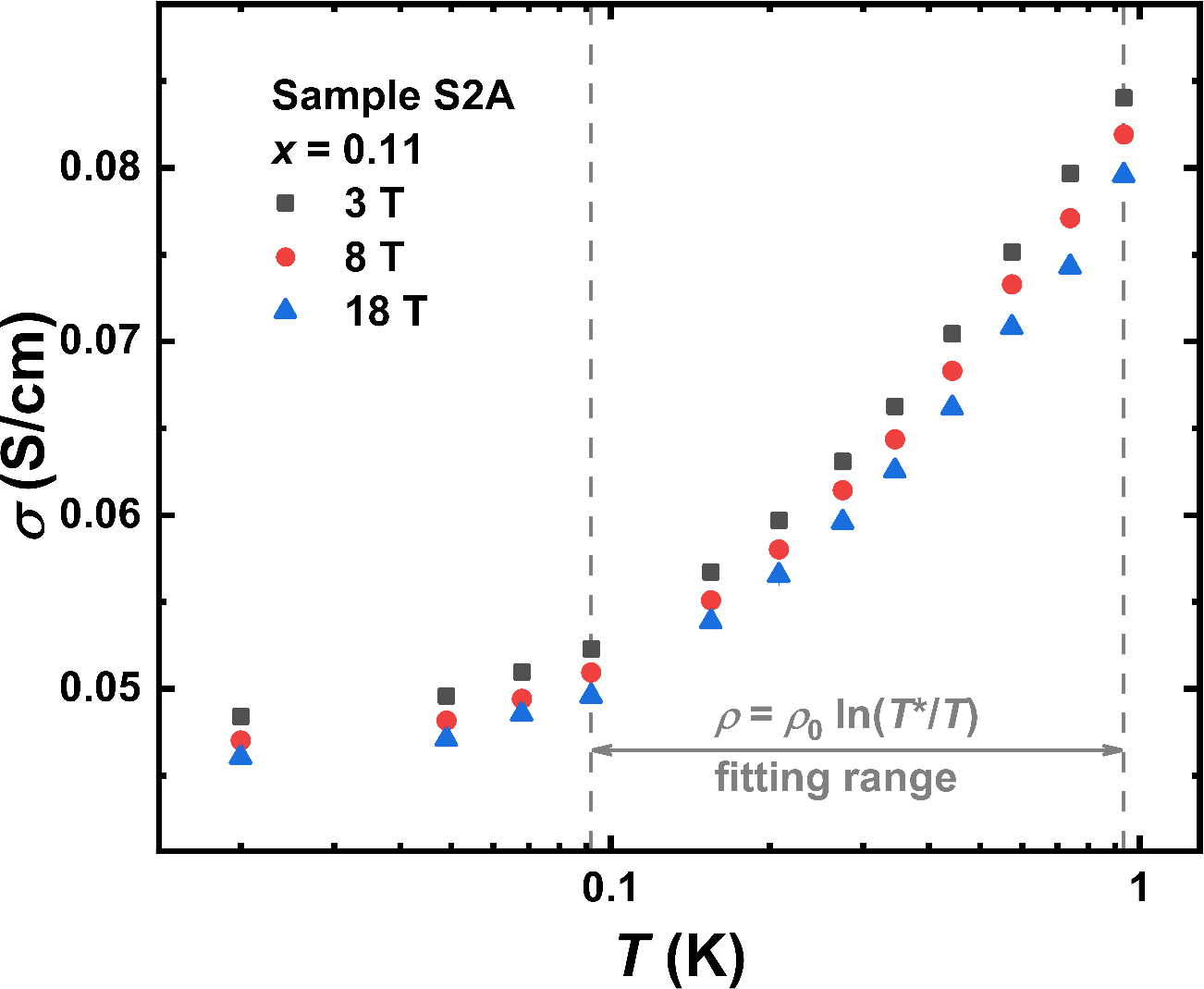}}
\caption{\textbf{Temperature dependence of the conductivity at high fields for sample S2A.}  SC is fully suppressed at $H> 2.5$~T (see main text).  The vertical dashed lines show a range of $T$ where $\rho=1/\sigma\propto\ln(1/T)$ is observed.
}
\label{fig:lnT_sigma}
\end{figure}
%


%
\begin{figure}[!tb]
\centerline{\includegraphics[width=\textwidth]{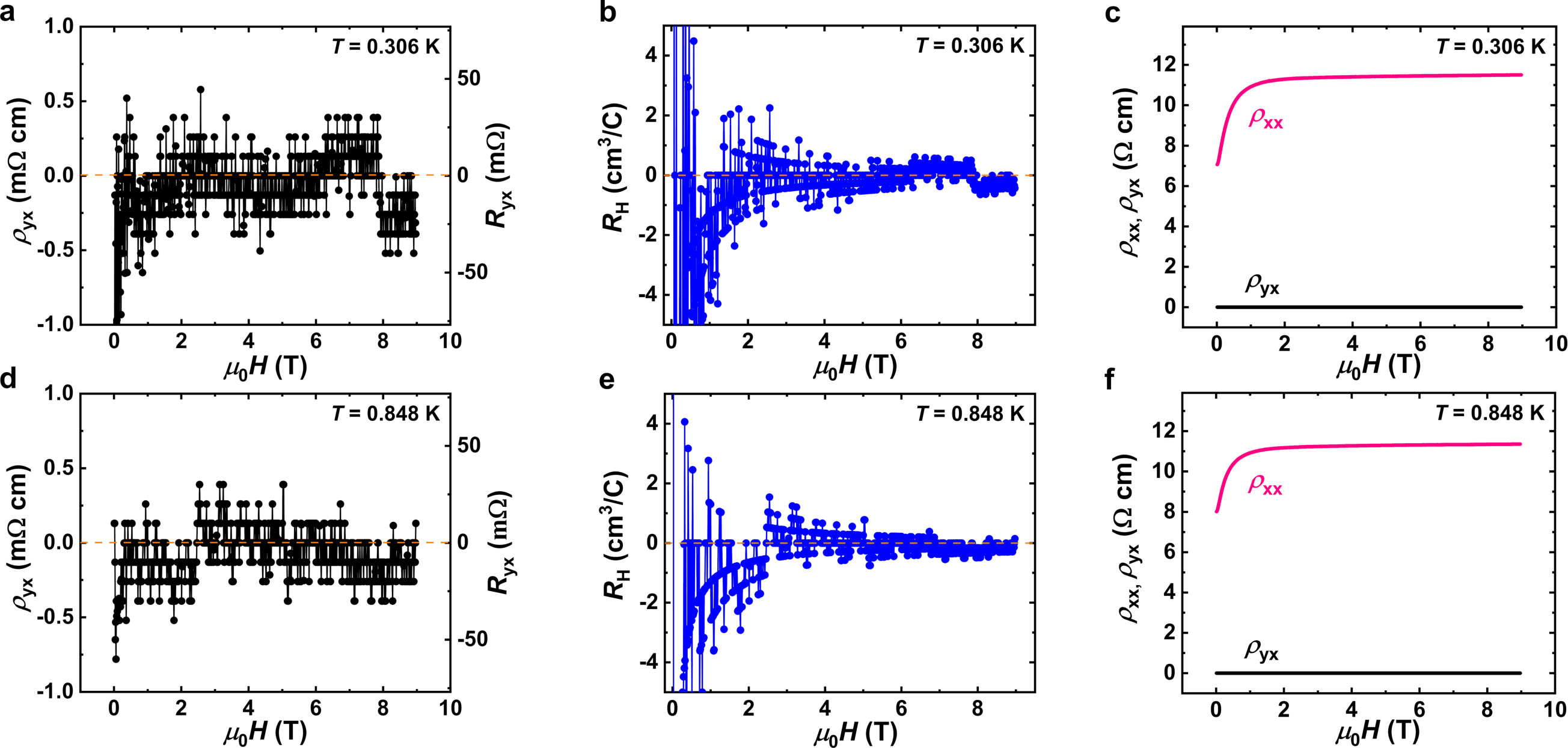}}
\caption{\textbf{Hall effect measurements in sample S2 ($\bm{x=0.11}$).} \textbf{a, b} The Hall resistivity $\rho_\mathrm{yx}(H)$ (resistance $R_\mathrm{yx}(H)$) and the corresponding Hall coefficient $R_\mathrm{H}= \rho_{\mathrm{yx}}(H)/H$, respectively, at $T=0.306$~K.  Dashed lines are guides to the eye.  \textbf{c} Comparison of the longitudinal resistivity $\rho_\mathrm{xx}(H)\equiv\rho(H)$ and $\rho_\mathrm{yx}(H)$ at $T=0.306$~K.   \textbf{d, e, f} Analogous results obtained at $T=0.848$~K.  At both $T$, $\rho_\mathrm{yx}(H)\approx 0$ at all fields.  In the normal state at high $H> 2.5$~T, $\rho_\mathrm{xx}\gg\rho_\mathrm{yx}$, such that $\rho_\mathrm{yx}$ is at least three orders of magnitude lower than $\rho_\mathrm{xx}$.}
\label{fig:Hall}
\end{figure}
%
\clearpage

%
\begin{figure}
\centerline{\includegraphics[width=0.9\textwidth]{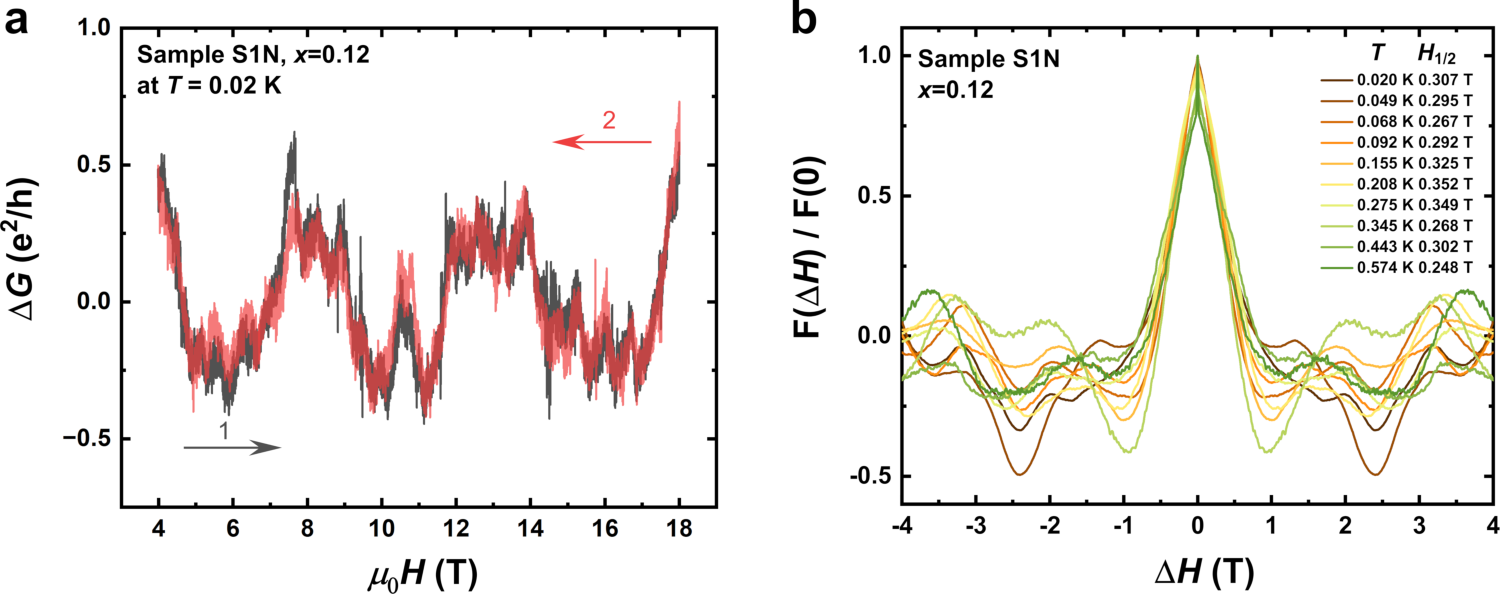}}
\caption{\textbf{Conductance fluctuations in the field-revealed normal state.} 
\textbf{a} Reproducible conductance fluctuations $\Delta G(H)= [1/R(H)] - \langle G(H)\rangle$, where $\langle G(H)\rangle$ is the background obtained by third-order polynomial fitting for $4$~T~$\leq H\leq 18$~T.  The arrows indicate the direction of consecutive field sweeps.  These data were taken several days before the data shown in Fig.~4c.  \textbf{b} The autocorrelation function $F(\Delta H)$ (see Supplementary Note 1) of the conductance fluctuations $\Delta G(H)$ shown in Fig.~4c for different $T$, with the corresponding characteristic fields $H_\mathrm{1/2}(T)$; $H_\mathrm{1/2}$ is determined as the half-width at half-maximum of $F(\Delta H)/F(0)$.
}
\label{fig:UCF_supp}
\end{figure}
%

\noindent{\large{\textbf{Supplementary Tables}}}

\begin{table*}[h!]
\centering
\caption{Scaling analysis of the measured sample resistivity $\rho(H,T)$ (see Supplementary Fig.~\ref{fig:rho_scaling_supp}).  For sample S1N, two attempts at scaling were made with  different nominal critical fields ``$H_\mathrm{c}$''.  The obtained values of $z\nu$ differ from sample to sample and even on both sides of the ``transition'' (sample S2B), indicating that the scaling analysis of the sample resistivity $\rho$ is not reliable.\\}
 \begin{tabular}{|c| c c c|} 
 \hline
  Sample & ``$H_\textrm{c}$''& $z\nu$ for $H<$``$H_\textrm{c}$''&$z\nu$ for $H>$``$H_\textrm{c}$''  \\ [0.5ex] 
 \hline
S1N, attempt\#1&$0.485~\textrm{T}$ &$1.73 \pm 0.01$
&$1.74 \pm 0.01$\\
S1N, attempt\#2&$0.525~\textrm{T}$ &not a power law
&not a power law\\
\hline
S1&$0.855 \textrm{T}$& $0.86 \pm 0.02$&$0.86 \pm 0.01$\\
 \hline
S2A&$0.450 \textrm{T}$&$1.81 \pm 0.01$&$1.82 \pm 0.01$\\
 \hline
S2B&$0.635 \textrm{T}$&$0.93 \pm 0.01$&$1.48 \pm 0.01$\\ 
 \hline
 \end{tabular}
\label{Tab:rho scaling}
\end{table*}
\vspace*{12pt}

\clearpage
\noindent{\large{\textbf{Supplementary Note 1}}}

\noindent\textbf{Reproducible conductance fluctuations} 

\noindent 
The reproducible conductance fluctuations observed in \stf\, in the phase coexistence region can be understood largely within the framework of universal conductance fluctuations (UCFs) [S3, S4].  
Supplementary Fig.~\ref{fig:UCF_supp}b shows the autocorrelation function $F(\Delta H)=\langle \Delta G(H)\Delta G(H+\Delta H)\rangle$ of the conductance fluctuations $\Delta G (H)= [1/R(H)] - \langle G(H)\rangle$ from Fig.~4c. 
To compare fluctuations at all measurement $T$, $\langle G(H)\rangle$ was obtained by third-order polynomial fitting for $4$~T~$\leq H\leq 18$~T.  The characteristic correlation field $H_\mathrm{1/2}$ is determined as the half-width at half-maximum of $F(\Delta H)/F(0)$.  We find that $H_\mathrm{1/2}\lesssim 0.5$~T at all $T$ (Supplementary Fig.~\ref{fig:UCF_supp}b), but its precise temperature dependence could not be determined due to the error introduced by the choice of polynomial fitting for the monotonic background $\langle G(H)\rangle$ [S5].  
The correlation field $H_\mathrm{1/2}$ is set by the field at which the flux going through a phase coherent area changes by one flux quantum [S4], 
largely randomizing the relative phases of different trajectories.  This field scale thus determines the characteristic length $L_\mathrm{1/2}\sim\sqrt{(h/e)/H_\mathrm{1/2}}$, where $h$ is Planck's constant and $e$ is electron charge.  We find that $L_\mathrm{1/2}\sim 100$~nm at our measurement temperatures.  

In general, the correlation field $H_\mathrm{1/2}$ is determined by the smaller of the phase coherence length $L_\mathrm{in}$ and the thermal length $L_\mathrm{T}$, namely $L_\mathrm{1/2}=\mathrm{min}(L_\mathrm{in},L_\mathrm{T})$ [S4].  
Here we show that, at least at the lowest temperatures, $L_\mathrm{T}\gg 100$~nm, and hence the phase coherence length  $L_\mathrm{in}\sim100~\mathrm{nm}$.  The thermal length 
$L_\mathrm{T}=\sqrt{\frac{hD}{k_\mathrm{B}T}}$, where $D$ is the charge carrier diffusion constant and $k_\mathrm{B}$ is the Boltzmann constant. In 2D metallic systems, $D=\frac{1}{2}v_\mathrm{F}l=\frac{\sigma m^*}{2ne^2}v^2_\mathrm{F}$, where $v_\mathrm{F}$ is the Fermi velocity, $l=\frac{\sigma m^*}{ne^2}v_\mathrm{F}$ is the mean free path (elastic scattering length) determined from the Drude model, $\sigma$ is the conductivity, $m^*$ is the effective mass of the charge carrier, and $n$ is the charge carrier density (see, e.g., [S6]).  
For \stf~in its metallic phase, $\sigma\sim10~\mathrm{kS}/\mathrm{cm}$ [S7], 
$m^*\geq 5~m_\mathrm{e}$ (where $m_\mathrm{e}$ is the bare electron mass) for $x<0.28$ [S8], 
$n\sim 5.8\times10^{-4}$~\AA$^{-3}$ estimated from the lattice constant [S9], 
and the Fermi velocity is presumably similar to that of other organic charge transfer salts, $v_\mathrm{F}\sim5\times10^4~\mathrm{m/s}$ [S10, S11].
With these values, we obtain $L_T\gtrsim 920$~nm at $0.02$~K, i.e. about one order of magnitude higher than $L_\mathrm{1/2}\sim 100$~nm.  

For conductance fluctuations shown in Fig.~4c,~d, the root-mean-squared amplitude $\Delta G_\mathrm{rms}\sim 0.1~e^2/h$ at the lowest $T$.  Since $L_\mathrm{in}\ll L_\mathrm{T}$, thermal averaging can be neglected.  In that case, the suppression of $\Delta G_\mathrm{rms}$ below its universal, $\sim e^2/h$ value can be explained by the classical self-averaging for the given geometry of the samples.  Indeed, in each conducting layer with the edge length $L_c$ (along the $c$-axis, the direction of the current) and $L_b$ (along the $b$-axis), the wave function is phase-coherent within a square of size $L_\mathrm{in}\times L_\mathrm{in}$, in which $\Delta G_\mathrm{rms}\sim  {e^2}/{h}$. For a sample with $N$ uncorrelated conducting layers, 
$\Delta G_\mathrm{rms}\sim (e^2/h)\sqrt{(L_bL_\mathrm{in}^2)/{L_c^3}}\sqrt{N}$ [S4].  
For the geometry of our samples  (see Methods), $N\sim 10^5$, and thus $L_\mathrm{in}\sim100~\mathrm{nm}$ corresponds to $\Delta G_\mathrm{rms}\sim 0.1~e^2/h$, consistent with our observations.  The same classical self-averaging implies, though, that the conductance of each phase-coherent square is $\sim 10^{-3} e^2/h$ (since sample conductance $G=1/R$, with $R\sim 500~\Omega$ in the field-revealed normal state; see Fig.~2), whereas the theory for UCFs [S3, S4]
was formulated for homogeneous metals with conductance $\gg e^2/h$.  This apparent discrepancy is attributed to the presence of large-scale inhomogeneity in the phase coexistence region, e.g., insulating domains larger than phase-coherent squares, consistent with the Imry-Ma argument [S12] 
for the presence of domains of all sizes in disordered 2D systems.  Indeed, although there have been no studies of domain sizes in \stf, domains ranging from $\sim 10$~nm to $\sim10~\mu$m have been reported [S13-S16]
in other quasi-2D organics.  

\vspace*{12pt}

\clearpage

\noindent{\large{\textbf{Supplementary References}}}




\noindent [S1] Saito, Y., L{\"o}hle, A., Kawamoto, A., Pustogow, A. \& Dressel, M. Pressure-tuned superconducting dome in chemically-substituted {$\kappa$-(BEDT\--TTF)$_2$\-Cu$_2$(CN)$_3$}. \textit{Crystals} \textbf{11}, 817 (2021).\\
%
\noindent [S2] Campbell, A. M., Blunt, F. J., Johnson, J.D. \& Freeman, P. A. Quantitative determination of percentage superconductor in a new compound. \textit{Cryogenics} \textbf{31}, 732--737 (1991).\\
%
\noindent [S3] Lee, P. A. \& Stone, A. D.  Universal Conductance Fluctuations in Metals.  \textit{Phys. Rev. Lett.} {\bf 55}, 1622--1625 (1985).\\
%
\noindent [S4] Lee, P. A., Stone, A. D. \& Fukuyama, H. Universal conductance fluctuations in metals: Effects of finite temperature, interactions, and magnetic field. \textit{Phys. Rev. B} \textbf{35}, 1039--1070 (1987).\\
%
\noindent [S5] Lundeberg, M. B., Renard, J. \& Folk, J. A.  Conductance fluctuations in quasi-two-dimensional systems: A practical view. \textit{Phys. Rev. B} \textbf{86}, 205413 (2012).\\
%
\noindent [S6] Girvin, S. \& Yang, K. \textit{Modern Condensed Matter Physics} (Cambridge University Press: 2019).\\
%
\noindent [S7] Pustogow, A., Saito, Y., L{\"o}hle, A., Sanz Alonso, M., Kawamoto, A., Dobrosavljevi\'{c}, V., Dressel, M. \& Fratini, S. Rise and fall of Landau's quasiparticles while approaching the Mott transition. \textit{Nat. Commun.} \textbf{12}, 1571 (2021).\\
%
\noindent [S8] Pustogow, A. \textit{Unveiling Electronic Correlations in Layered Molecular Conductors by Optical Spectroscopy} Doctoral dissertation, University of Stuttgart (2017).\\
%
\noindent [S9] Saito, Y., R{\"o}sslhuber, R., L{\"o}hle, A., Sanz Alonso, M., Wenzel, M., Kawamoto, A., Pustogow, A. \& Dressel, M. Chemical tuning of molecular quantum materials {$\kappa$-(BEDT\--TTF)$_2$\-Cu$_2$(CN)$_3$}: from the Mott-insulating quantum spin liquid to metallic Fermi liquid. \textit{J. Mater. Chem. C} \textbf{9}, 10841--10850 (2021).\\
%
\noindent [S10] Kovalev, A. E., Hill, S. \& Qualls, J. S. Determination of the Fermi velocity by angle-dependent periodic orbit resonance measurements in the organic conductor $\alpha$-(BEDT-TTF)$_2$KHg(SCN)$_4$. \textit{Phys. Rev. B} \textbf{66}, 134513 (2002).\\
%
\noindent [S11] Kimata, M., Oshima, Y., Ohta, H., Koyama, K., Motokawa, M., Yamamoto, H. M. \& Kato, R. Magnetooptical measurements of $\beta$''-(BEDT-TTF)(TCNQ). \textit{Physica B} \textbf{346-347}, 382-386 (2004).\\
%
\noindent [S12] Imry, Y. \& Ma, S.-k. Random-field instability of the ordered state of continuous symmetry. \textit{Phys. Rev. Lett.} \textbf{35}, 1399--1401 (1975).\\
%
\noindent [S13] Sasaki, T., Yoneyama, N., Kobayashi, N., Ikemoto, Y. \& Kimura, H. Imaging phase separation near the Mott boundary of the correlated organic superconductors {$\kappa$-(BEDT-TTF)$_2$X}. \textit{Phys. Rev. Lett.} \textbf{92}, 227001 (2004).\\
%
\noindent [S14] Sasaki, T., Yoneyama, N., Suzuki, A., Kobayashi, N., Ikemoto, Y. \& Kimura, H. Real space imaging of the metal-insulator phase separation in the band width controlled organic Mott system {$\kappa$-(BEDT-TTF)$_2$Cu[N(CN)$_2$]Br}. \textit{J. Phys. Soc. Jpn.} \textbf{74}, 2351--2360 (2005).\\
%
\noindent [S15] M\"{u}ller, J., Brandenburg, J. \& Schlueter, J. A. Magnetic-field induced crossover of superconducting percolation regimes in the layered organic Mott system {$\kappa$-(BEDT-TTF)$_2$Cu[N(CN)$_2$]Cl}. \textit{Phys. Rev. Lett.} \textbf{102}, 047004 (2009).\\
%
\noindent [S16] Pustogow, A., McLeod, A. S., Saito, Y., Basov, D. N. \& Dressel, M.  Internal strain tunes electronic correlations on the nanoscale. \textit{Sci. Adv.} \textbf{4}, eaau9123 (2018).


\end{document}